%% file: triple.tex
\documentclass[aps,showpacs,showkeys,11pt]{revtex4}

\usepackage{graphicx,epic,eepic,epsfig,amsmath,latexsym,amssymb,verbatim}

\usepackage{theorem}
\newtheorem{definition}{Definition}[section]

\newtheorem{lemma}[definition]{Lemma}

\newtheorem{theorem}[definition]{Theorem}

\def\squareforqed{\hbox{\rlap{$\sqcap$}$\sqcup$}}
\def\qed{\ifmmode\squareforqed\else{\unskip\nobreak\hfil
\penalty50\hskip1em\null\nobreak\hfil\squareforqed
\parfillskip=0pt\finalhyphendemerits=0\endgraf}\fi}
\def\endenv{\ifmmode\;\else{\unskip\nobreak\hfil
\penalty50\hskip1em\null\nobreak\hfil\;
\parfillskip=0pt\finalhyphendemerits=0\endgraf}\fi}
\newenvironment{proof}{\noindent \textbf{{Proof~} }}{\qed}

\newcommand{\exampleTitle}[1]{\textbf{(#1)}}

\newcommand{\proofComment}[1]{\exampleTitle{#1}}

\mathchardef\ordinarycolon\mathcode`\:
\mathcode`\:=\string"8000
\def\vcentcolon{\mathrel{\mathop\ordinarycolon}}
\begingroup \catcode`\:=\active
  \lowercase{\endgroup
  \let :\vcentcolon
  }

\newcommand{\nc}{\newcommand}
\nc{\rnc}{\renewcommand}
\nc{\beq}{\begin{equation}}
\nc{\eeq}{{\end{equation}}}
\nc{\beqa}{\begin{eqnarray}}
\nc{\eeqa}{\end{eqnarray}}
\nc{\lbar}[1]{\overline{#1}}
\nc{\bra}[1]{\langle#1|}
\nc{\ket}[1]{|#1\rangle}
\nc{\ketbra}[2]{|#1\rangle\!\langle#2|}
\nc{\braket}[2]{\langle#1|#2\rangle}
\nc{\proj}[1]{| #1\rangle\!\langle #1 |}
\nc{\avg}[1]{\langle#1\rangle}
\rnc{\max}{\operatorname{max}}
\nc{\Rank}{\operatorname{Rank}}
\nc{\smfrac}[2]{\mbox{$\frac{#1}{#2}$}}
\nc{\Tr}{\operatorname{Tr}}
\nc{\ox}{\otimes}
\nc{\dg}{\dagger}
\nc{\dn}{\downarrow}
\nc{\cA}{{\cal A}}
\nc{\cB}{{\cal B}}
\nc{\cC}{{\cal C}}
\nc{\cD}{{\cal D}}
\nc{\cE}{{\cal E}}
\nc{\cF}{{\cal F}}
\nc{\cG}{{\cal G}}
\nc{\cH}{{\cal H}}
\nc{\cI}{{\cal I}}
\nc{\cJ}{{\cal J}}
\nc{\cK}{{\cal K}}
\nc{\cL}{{\cal L}}
\nc{\cM}{{\cal M}}
\nc{\cO}{{\cal O}}
\nc{\cP}{{\cal P}}
\nc{\cR}{{\cal R}}
\nc{\cS}{{\cal S}}
\nc{\cT}{{\cal T}}
\nc{\cX}{{\cal X}}
\nc{\cZ}{{\cal Z}}
\nc{\optimal}{^*}
\nc{\csupp}{{\operatorname{csupp}}}
\nc{\qsupp}{{\operatorname{qsupp}}}
\nc{\esupp}{{\operatorname{esupp}}}
\nc{\var}{\operatorname{var}}
\nc{\rar}{\rightarrow}
\nc{\lrar}{\longrightarrow}
\nc{\polylog}{\operatorname{polylog}}

\def\a{\alpha}

\def\d{\delta}
\def\e{\epsilon}

\def\l{\lambda}

\def\r{\rho}

\def\ph{\varphi}

\def\o{\omega}

\def\D{\Delta}

\nc{\RR}{{{\mathbb R}}}
\nc{\CC}{{{\mathbb C}}}
\nc{\FF}{{{\mathbb F}}}
\nc{\NN}{{{\mathbb N}}}
\nc{\ZZ}{{{\mathbb Z}}}
\nc{\PP}{{{\mathbb P}}}
\nc{\QQ}{{{\mathbb Q}}}
\nc{\UU}{{{\mathbb U}}}
\nc{\EE}{{{\mathbb E}}}
\nc{\id}{{\mathbb I}}

\begin{document}

\title{{\Large Generalized remote state preparation:}\\
Trading cbits, qubits and ebits in quantum communication}

\author{Anura Abeyesinghe}
\email{anura@caltech.edu}
\affiliation{
Institute for Quantum Information, 
Physics Department,
Caltech 103-33, 
Pasadena, CA 91125, USA}
\author{Patrick Hayden}
\email{patrick@cs.caltech.edu}
\affiliation{
Institute for Quantum Information, 
Physics Department,
Caltech 103-33, 
Pasadena, CA 91125, USA}

\date{\today}

\begin{abstract}
We consider the problem of communicating quantum states by simultaneously
making use of a noiseless classical channel, a noiseless quantum channel
and shared entanglement. We specifically study the version of the problem 
in which the sender is given knowledge of the state to be communicated.
In this setting, a trade-off arises between the three resources, some portions
of which have been investigated previously in the contexts of 
the quantum-classical trade-off in data compression, remote state
preparation and superdense coding of quantum states, each of which
amounts to allowing just two out of these three resources.
We present a formula for the triple resource trade-off that reduces its
calculation to evaluating the data compression trade-off formula.
In the process, we also construct protocols achieving all the optimal
points. These turn out to be achievable by trade-off coding and
suitable time-sharing between
optimal protocols for cases involving two resources out of the three 
mentioned above.
\end{abstract}

\pacs{03.65.Ta, 03.67.Hk}

\keywords{compression, superdense coding, remote state preparation, 
entanglement}

\maketitle

\section{Introduction}
Quantum information theory can be described as the effort to identify and
quantify the basic resources required to communicate or, more generally,
process information in a quantum mechanical setting. 
The dual goals of identifying new protocols and demonstrating 
their optimality have, respectively, helped to expose the surprising range of
information processing tasks facilitated by quantum mechanics
and highlighted the subtle ways in which physics dictates limitations
on the transmission and processing of information.

Part of the appeal of the information theoretic paradigm is that it emphasizes
the notions of interconvertibility and simulation. Identifying basic
resources and evaluating their interconvertibility provides a general
strategy for systematically charting the capabilities of 
quantum mechanical systems.
Some early successes of this approach include Schumacher's quantum noiseless
coding theorem~\cite{S95,OP93}, 
which demonstrated that a single number quantifies the
compressibility of memoryless sources of quantum states, and the
theory of pure state bipartite entanglement, where a single number, likewise,
determines the asymptotic interconvertibility of entanglement~\cite{BBPS96}. 
More
recently, we have seen how to evaluate the interconvertibility of quantum
memories~\cite{K02} 
and even seen that the rate at which one noisy quantum channel
can simulate any other (in the presence of entanglement and with certain
restrictions on the input) is controlled again
by a single number, the channel's entanglement-assisted capacity~\cite{QRST}.

From the point of view of communication theory, these results identify
three basic and inequivalent resources: noiseless classical
channels, noiseless quantum channels and maximally entangled states.
Other inequivalent resources exist, of course. One such,
classically correlated bits, will prove useless for the 
problem we investigate. Noisy versions of the basic list of 
three resources identified above potentially adds many others but 
we don't study them here. Those caveats aside, the three basic resources
serve as formalized versions of abstract ``classicality'', ``quantumness'' 
and ``nonlocality'', quantifiable in units of classical bits (cbits),
quantum bits (qubits) and maximally entangled qubits (ebits). While
the three basic resources are inequivalent, relationships exist between
them. Because cbits can be encoded in qubits and ebits can
be established by sending qubits, the noiseless quantum channel is
(in this narrow sense) the strongest of the three. Because it is impossible
to establish entanglement using classical communication or to communicate
using only entanglement, ebits and cbits are simply incomparable; neither
is truly stronger than the other.

In the present work, we quantify the relationship between the three 
resources for a basic task in quantum information 
theory: communicating quantum states from a sender to a receiver
(and, more generally, sharing entangled states between them).
There are at least two variations on the task, depending on whether or not 
the sender has knowledge of the states she is required to communicate.
If she is only given a copy of the quantum state and not a description, 
we describe the source as hidden and the encoding as oblivious (or blind).
At the
other extreme, if she is told which state she is required to transmit,
we describe the source as visible and the encoding as non-oblivious.
(Sometimes in the quantum information literature the adjective 
``visible'' is also applied, somewhat nonsensically, to the encoding.)
While the distinction
makes no difference in classical information theory, quantum mechanical
restrictions on the sender's ability to measure without causing a disturbance
lead to very different results for the two tasks in the quantum case.
(Compare, for example, the results of Refs.~\cite{BHJW01,KI02} 
and \cite{HJW02}.)
Our emphasis here is on the visible scenario since there is generically
only a trivial trade-off for the blind encoder case: 
using teleportation, two cbits
and one ebit can be used to simulate a noiseless one-qubit channel but
no other interesting trade-offs are possible.

In the visible scenario, the relationship between the three resources
becomes much more varied. When no quantum channel is permitted, we recover
the problem known as remote state preparation~\cite{Lo99,BDSSTW01}, 
while forbidding use of the
classical channel leads to superdense coding of quantum 
states~\cite{BW92,HHL03}. 
Likewise, if entanglement 
is not permitted, we recover the trade-off between classical and
quantum communication solved in Ref.~\cite{HJW02}. The present paper 
completely solves the problem of trading all three resources against each
other, finding that optimal protocols for any combination of resources
can be constructed by appropriate combinations of the protocols representing
the extremes identified above. Such a clean resolution
in terms of previously discovered building blocks is encouraging: it confirms
yet again the simplifying power of the resource-based approach, this
time yielding a manageable taxonomy of optimal protocols for the
triple trade-off problem.

The rest of the paper is structured as follows. Section \ref{sec:defn} 
defines the problem rigorously and describes previous results for
the cases when one of the three resources is not used, along with
some minor extensions.
Section \ref{connections} 
studies the relationship between the trade-off between qubits and cbits
in quantum data compression (QCT) and the trade-off between ebits and cbits
in remote state preparation (RSP).
In section \ref{main} these 
connections and the results described in section \ref{sec:defn} are 
used to obtain optimal protocols and optimal resource trade-offs for 
communicating quantum states when all three resources are used simultaneously: 
the full ``triple trade-off''. 

We use the following conventions throughout the paper. If 
$\cE_{AB} = \{ \ph_i^{AB}, p_i \}$ is an ensemble of bipartite states
then we write $\cE_A$ for the ensemble $\{ \ph_i^A, p_i \}$ of reduced
states on system $A$.
Sometimes we omit subscripts (or superscripts) labelling subsystems,
in which case the largest subsystem on which the ensemble (or state)
has been defined should be assumed: $\cE = \cE_{AB}$ and 
$\ph_i = \ph_i^{AB}$.
We identify states with their density operators and if $\ket{\ph}$ is a 
pure state, we use the notation $\ph = \proj{\ph}$ for its density operator.
The function $S(\rho)$ is the von Neumann entropy 
$S(\rho) = -\Tr \rho \log \rho$ and $S(\cE)$ the von Neumann entropy
of the average state of the ensemble $\cE$. Functions like $S(A|B)_\r$ and
$S(A:B|C)_\r$ are defined in the same way as their classical counterparts:
\begin{equation}
S(A:B|C)_\r = S(\r^{AC}) + S(\r^{BC}) - S(\r^{ABC}) - S(\r^C),
\end{equation} 
for example. $\chi(\cE)$ is
the Holevo $\chi$ quantity of $\cE$~\cite{H73}. Given a bipartite ensemble 
$\cE_{AB} = \{ \ph_i^{AB}, p_i \}$,
we also make use the abbreviations $S=S(\cE_{B})$, 
$\bar{S} = \sum_i p_i \ph_i^B$, $\chi=\chi(\cE_{B})$ and $H=H(p_{i})$. 
Throughout, $\log$ and $\exp$ are taken base $2$.

\section{Definition of the problem and previous results} \label{sec:defn}

\begin{figure}
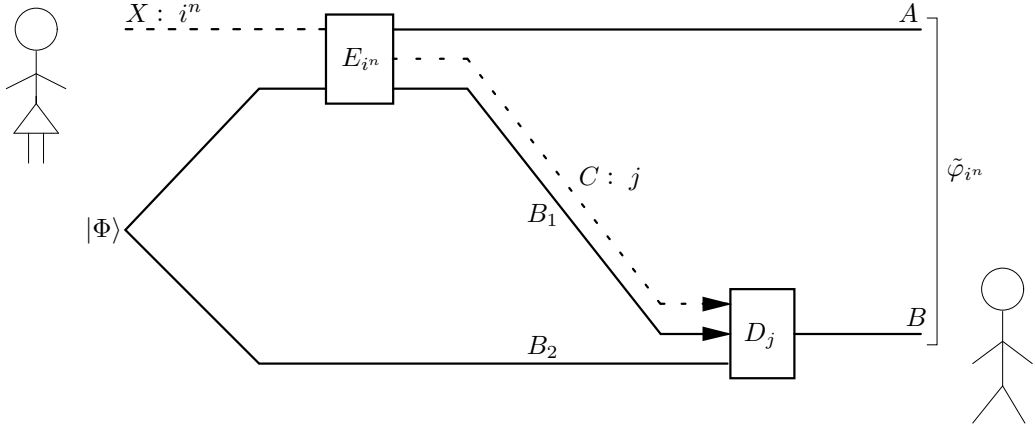

\begin{center}
\input circuit.eepic
\caption{In the above quantum circuit diagram for generalized remote 
state preparation time goes from left to right, solid lines represent 
quantum registers and dashed lines represent classical registers. The 
registers connected in the left represent a maximally entangled state of
$\log d_E$ ebits initially shared between Alice and Bob. The $\log d_Q$-qubit 
quantum register $B_1$ is sent from Alice to Bob, as is the $\log d_C$ cbit
classical message $m$. Alice's encoding operation is 
denoted by $E_{{i}^{n}}$ and Bob's decoding operation, which is 
conditioned on $m$, by $D_m$.}
\label{fig:circuit}
\end{center}
\end{figure}

We now give a more formal definition of the task to be completed by the
sender and receiver, henceforth, respectively Alice and Bob. The reader
can also refer to figure \ref{fig:circuit}, which illustrates
the definition. We consider
an ensemble of bipartite quantum states 
$\cE = \{ \ket{\ph_i}^{AB}, p_i \}$ on a finite-dimensional Hilbert space
$\cH_{AB} = \cH_A \ox \cH_B$ and the product ensembles 
$\cE^{\ox n} = \{ \ket{\ph_{i^n}}^{AB}, p_{i^n} \}$ on $\cH_{AB}^{\ox n}$, 
where
\begin{eqnarray*}
i^n &=& i_1 i_2 \dots i_n, \\
p_{i^n} &=& p_{i_1} p_{i_2} \dots p_{i_n} \quad \mbox{and} \\
\ket{\ph_{i^n}} &=& 
	\ket{\ph_{i_1}} \ox \ket{\ph_{i_2}} \ox \dots \ox \ket{\ph_{i_n}}.
\end{eqnarray*}
At the end of the protocol, Alice and Bob are to reproduce the states
of the bipartite ensemble with high fidelity. (Regardless of whether
pure states are \emph{prepared} in Bob's system, or entangled states
are \emph{shared} between Alice and Bob, we will always refer to the
task simply as \emph{communicating} from Alice to Bob.)
We imagine that there is a noiseless classical channel from
Alice to Bob capable of sending one of $d_C$ messages, a noiseless quantum
channel capable of sending a $d_Q$-dimensional quantum system and a 
maximally entangled state 
$\ket{\Phi} = d_E^{-1/2} \sum_{i=1}^{d_E} \ket{i}\ket{i}$ 
of Schmidt rank $d_E$.
A source provides Alice with $i^n$, drawn with probability
$p_{i^n}$, at which point Alice applies a quantum operation $E_{i^n}$ to her
half of $\ket{\Phi}$ that without loss of generality has output of the form
\begin{equation}
\sum_{j=1}^{d_C} \r^{A B_1 B_2}_{i^n,j} \ox q(j|i^n) \proj{j}^C,
\end{equation}
where $B_1$ is a $d_Q$-dimensional quantum system, $B_2$ is the
quantum system supporting Bob's half of $\ket{\Phi}$,
the states $\{\ket{j}\}$ are orthonormal (\emph{i.e.} classical) 
and $q(\cdot | i)$ is a probability 
distribution. Alice then sends register $B_1$ to Bob over her
noiseless quantum channel and $C$ to Bob over the noiseless classical
channel. The protocol is completed by Bob performing a quantum operation
$D_j$ on registers $B_1$ and $B_2$. Write $\tilde{\ph}_{i^n}$ for the
joint Alice-Bob output state averaged over different values of $j$.
We say that the protocol has fidelity $1-\e$ if
\begin{equation}
\sum_{i^n} p_{i^n} \bra{\ph_{i^n}} \tilde{\ph}_{i^n} \ket{\ph_{i^n}}
\geq 1 - \e.
\end{equation}
Likewise, $(R,Q,E)$ is an achievable rate triple for the ensemble $\cE$
if for all $\d,\e > 0$ there exists $N$ such that for all $n > N$ there 
is a protocol for $\cE^{\ox n}$ with fidelity $1 - \e$ and
\begin{eqnarray}
\frac{1}{n} \log d_C \leq R + \d \quad
\frac{1}{n} \log d_Q \leq Q + \d \quad
\frac{1}{n} \log d_E \leq E + \d.
\end{eqnarray}

Our goal will be to identify these achievable triples. In particular,
we will find a formula for the function
\begin{equation}
E^*(R,Q) = \inf\{ E : (R,Q,E) \; \mbox{is achievable} \}.
\end{equation}

We refer to  rate triples of the form $(R,Q,E^{*}(R,Q))$ as optimal rate 
triples and the protocols that achieve them as optimal protocols. 
We will indicate that a rate triple $(R,Q,E)$ is optimal by writing
it as $(R,Q,E)\optimal$.
Throughout the paper, unless otherwise stated, 
all entropic quantities will be taken
with respect to 4-partite states $\o$ of the following form:
\begin{equation} \label{eqn:omega} 
\o = \sum_i p_i \proj{i}^X \ox \ph_i^{AB} \ox 
	\sum_{j=1}^{m+1} p(j|i) \proj{j}^C,
\end{equation}
where $m$ is the number of states in $\cE_{AB}$ (if that number is finite),
and $p(\cdot|\cdot)$ is a classical noisy channel.
Note that for all such states 
\begin{equation}\label{eqn:entomega}
S(X:B|C) = S(B|C)- \bar{S}, \quad \mbox{where}\; 
	\bar{S}=\sum_{i}p_{i}S(\ph_{i}^{B}),
\end{equation}
a fact that will be useful later.

Before moving on to the general problem, we consider the special 
cases given by setting one of the three rates to zero.

\subsection{$Q=0:$ Remote state preparation (RSP)} \label{sec:thmE}

This problem was studied extensively in Ref.~\cite{BHLSW03}. It is impossible
to achieve an entanglement rate of less than $\sum_i p_i \ph_i^B$, essentially
because
that is the amount of entanglement shared between Alice and Bob at the
end of any successful protocol. The optimal cbit rate when the entanglement
is minimal is just $H(p_i)$, meaning that the simple protocol consisting
of Alice communicating $i^n$ to Bob and then the pair performing entanglement
dilution is optimal. At the other extreme, the cbit rate is minimized 
(at least for irreducible sources) by 
a protocol achieving the rate $(\chi(\cE_B),0,S(\cE_B))$. 
In general, we introduce the function
\begin{equation} \label{eqn:defnE}
E^*(R) = \inf \{ E : (R,0,E) \;\mbox{is achievable} \}.
\end{equation}
This choice, a slight abuse of notation given our earlier definition
of a function $E^*$ with two arguments, is chosen for consistency with
the remote state preparation paper. Note that $E^*(R) = E^*(R,0)$.
We have the following theorem from Ref.~\cite{BHLSW03}:
\begin{theorem} \label{thm:RSP}
For the ensemble $\cE = \{\ket{\ph_i}^{AB},p_i\}$ of pure bipartite
states and $R \geq 0$,
\begin{equation} \label{eqn:thmE}
E^*(R) = \min\{S(B|C) : S(X:BC) \leq R\},
\end{equation}
where the entropic quantities are with respect to the state $\o$,
minimization is over all $4$-partite states $\o$ of the 
form of Eq.~(\ref{eqn:omega})
with classical channels $p(j|i)$, and $m$ the number of states in $\cE$.
$E^*$ is convex, continuous and strictly decreasing in the interval in
which it takes positive values.
\end{theorem}
We will also use the simple fact that the inequality in 
Eq.~(\ref{eqn:thmE}) can be replaced by equality.

\subsection{$E=0$: Quantum-classical trade-off (QCT)}

The case where the ensemble $\cE$ consists only of product states 
$\ket{\ph_i}^{AB} = \ket{0}^A\ket{\ph_i}^B$ was the focus of
Ref.~\cite{HJW02}. At the extreme when 
$R=0$, only quantum communication is permitted so the problem of finding
achievable rates is answered by the quantum noiseless coding theorem:
$(0,S(\cE_B),0)$ is an \emph{optimal point}, in the sense that none of the
three rates can be reduced. Likewise, the optimal point when $Q=0$
is given by $(H(p_i),0,0)$, meaning that Alice has no better strategy than
to communicate the label $i^n$ to Bob. More generally, when the ensemble
is allowed to contain entangled states, the techniques of 
Refs.~\cite{HJW02,BHLSW03} are easily adapted to yield a formula for
\begin{equation} \label{eqn:eqnE}
Q^*(R) = \inf\{Q : (R,Q,0) \;\mbox{is achievable} \}.
\end{equation}
In particular, we have the following analog of theorem \ref{thm:RSP}:
\begin{theorem} \label{thm:QCT}
For the ensemble $\cE = \{\ket{\ph_i}^{AB},p_i\}$ of pure bipartite
states and $R \geq 0$,
\begin{equation} \label{eqn:thmQ}
Q^*(R) = \min\{S(B|C) : S(X:C) \leq R\},
\end{equation}
where the entropic quantities are with respect to the state $\o$,
minimization is over all $4$-partite states $\o$ of the 
form of Eq.~(\ref{eqn:omega})
with classical channels $p(j|i)$, and $m$ the number of states in $\cE$.
$Q^*$ is convex, continuous and strictly decreasing in the interval in
which it takes positive values. There exists a critical value of $R$, 
hereafter referred to as $H_c$ such that $R+Q^{*}(R)=S(B)$ for $R \leq H_c$
and $R+Q^*(R) > S(B)$ otherwise.
\end{theorem}
As before, the inequality in Eq.~(\ref{eqn:thmQ}) can be replaced by equality.

\subsection{$R=0:$ Superdense coding of quantum states (SDC)}

Ref.~\cite{HHL03} showed that it is possible to communicate arbitrary
$d^2$-dimensional quantum states using $\log d + o(\log d)$ qubits,
$\log d + o(\log d)$ ebits and shared random bits. For exploring the
trade-off of quantum resources, we need a 
variation on this result that applies to ensembles of entangled states:
using his coherent classical communication technique, Harrow has shown
that
\begin{equation}
\left(0,\smfrac{1}{2} \chi(\cE_B), S(\cE_B) - \smfrac{1}{2} \chi(\cE_B) \right)
\end{equation}
is an achievable rate triple \cite{H03}. Using his construction,
we can easily find the $R=0$ trade-off curve:
\begin{theorem} \label{thm:sdc}
For the ensemble $\cE = \{\ket{\ph_i}^{AB},p_i\}$ of pure bipartite
states and $Q \geq 0$,
\begin{equation} \label{eqn:thmS}
E^*(0,Q) = \left\{
\begin{array}{cc}
S(\cE_B) - Q & \quad \mbox{if}\; Q \geq \chi(\cE_B) / 2 \\
+\infty      & \quad \mbox{otherwise}.
\end{array}
\right.
\end{equation}
\end{theorem}
\begin{proof}
Since $(0,S,0)$ and $(0,\chi/2,S-\chi/2)$ ($S$ and $\chi$ are defined
in the introduction) are both achievable rate triples,
any convex combination of the two is an achievable rate triple corresponding
to a time-shared protocol. Thus, if $0 \leq \l \leq 1$,
\begin{equation}
\left(0,\l S + (1-\l) \chi / 2, (1-\l)(S - \chi / 2) \right)
\end{equation}
is achievable. Suppose these points are not optimal. Then there
exists $\e > 0$ such that
\begin{equation}
\left(0,\l S + (1-\l) \chi / 2 , (1-\l)(S - \chi / 2) - \e \right)
\end{equation}
is optimal. By using quantum communication to establish entanglement,
however, protocols achieving this rate can be converted into protocols
with the rate triple
\begin{eqnarray}
\left(0,\l S + (1-\l) \chi / 2 
	+ (1-\l)(S - \chi / 2) - \e, 0 \right) 
= (0, S- \e, 0),
\end{eqnarray}
contradicting the optimality of Schumacher compression. We conclude
that $E^*(0,Q) = S - Q$ when this conversion is possible, that is,
when $Q \geq \chi / 2$. This condition is required by causality. (For
a detailed proof, see section \ref{subsec:forbidden}.)
\end{proof}
The simple argument used in the proof of theorem \ref{thm:sdc} is
characteristic of what will follow. Our evaluation of $E^*(R,Q)$ will
be accomplished via operational reductions to the three extremal
cases we have now completed, just as theorem \ref{thm:sdc} was demonstrated
using a reduction from the unknown $E^*(0,Q)$ curve to the known
Schumacher compression point.

Later we will also have occasion to make use of the following analog
of the QCT and RSP constructions.
Given a state $\o$ of the form of
Eq.~(\ref{eqn:omega}), the trade-off coding technique from Ref.~\cite{HJW02}
then gives protocols achieving all the rate triples of the form
\begin{equation} \label{eqn:superdensePoints}
\left(
S(X:C),\smfrac{1}{2}S(X:B|C), S(B|C) - \smfrac{1}{2}S(X:B|C)
\right).
\end{equation}
Briefly, once an optimal channel $p(j|i)$ is chosen, Alice and Bob can
share (typical) $j^n = j_1 \dots j_n$ at a cost of  
$nS(X:C)+o(n)$ bits of communication plus shared random bits using
the Reverse Shannon Theorem~\cite{BSST02}. 
Harrow's protocol is then used on the induced ``conditional'' ensembles
\begin{eqnarray}
\{ \ket{\ph_{i^n}}^{AB},q(i^n|j^n) &=& q(i_1|j_1)\dots q(i_n|j_n)\}, \quad 
	\mbox{where} \nonumber \\
q(i|j) &=& \left( \sum_{i'} p_{i'} p(j|i') \right)^{-1} p(j|i) p_i.
\end{eqnarray}
The shared random bits are then seen to be unnecessary because we only
require high fidelity on average (so that some particular value of the
shared random bits can be used). Evaluation of the rates for the approach
gives exactly Eq.~(\ref{eqn:superdensePoints}). 

Given any $(R,Q^{*}(R),0)$ there is a state $\o$ of the form 
Eq.~(\ref{eqn:omega}) for which $(S(X:C),S(B|C),0) =(R,Q^{*}(R),0)$. 
For this state, we therefore find a new achievable rate triple:
\begin{equation} \label{eqn:SDC}
 \left(S(X:C),\smfrac{1}{2}S(X:B|C), S(B|C) - \smfrac{1}{2}S(X:B|C)\right)= \left( R , \smfrac{1}{2}(Q^{*}(R)-\bar{S}) , \smfrac{1}{2}(Q^{*}(R)+\bar{S} ) \right),
\end{equation}
where we have used Eq.~(\ref{eqn:entomega}) to arrive at the 
expression on the right hand side.

\section{Relating optimal QCT and optimal RSP} \label{connections}

Any protocol for quantum-classical compression can be converted into
an RSP protocol by using RSP to send the compressed qubits. One might
hope that if the original QCT point was optimal that the resulting RSP
point would also be optimal. For classical rates above $H_c$ this is indeed the
case but otherwise it need not be. Consider, for example, the ensemble 
consisting of the orthonormal states $\ket{0}$ and $\ket{1}$, each
occurring with probability $1/2$. In this case, $Q^*(0) = 1$ but the
corresponding RSP protocol would wastefully consume $1$ cbit and $1$
ebit per signal when $1$ cbit and no entanglement are sufficient.

As an aside, while we have described a natural way to convert optimal
QCT protocols into optimal RSP protocols (that works when $R \geq H_c$),
there is no known way to do the opposite. An appendix to Ref.~\cite{BHLSW03},
however, demonstrates the existence of 
just such an operational reduction but only
under the assumption that the mixed state compression conjecture is 
true. (See Refs.~\cite{H00,BCFJS01,W02} for more details on the
conjecture.)

The following two lemmas formally express the relationship between
optimal QCT and optimal RSP:

\begin{lemma} \label{lem:1}
When $R \geq H_c$, $E^{*}(R+Q^{*}(R)-\bar{S})= Q^{*}(R)$. Otherwise,
$E^*(R+Q^*(R)-\bar{S}) = Q^*(H_c)$.
\end{lemma}
\begin{proof}
We begin by showing that $E^{*}(R+Q^{*}(R)-\bar{S})\leq Q^{*}(R)$.  
We know that $(S(X:BC),0,S(B|C))$ is an achievable rate triple for any $ \o $ of the form of Eq.~(\ref{eqn:omega}). In particular, it is achievable when
$\left(S(X:C), S(B|C), 0\right) = \left(R, Q^{*}(R), 0 \right)$,
in which case
\begin{eqnarray}
            \left( S(X:BC),0,S(B|C)\right) & = & \left(S(X:C)+S(B|C)-\bar{S},0,S(B|C)\right)\\
                                 & = & \left(R+Q^{*}(R)-\bar{S},0,Q^{*}(R) \right).
 \end{eqnarray}
This proves the claim. Note that this inequality is true regardless of whether
$R$ is greater or less than $H_c$.

We now prove the opposite inequality:
$E^{*}(R+Q^{*}(R)-\bar{S}) \geq Q^{*}(R)$ when $R \geq H_c$. 
Substituting our expressions for $E^{*}(R)$ and $Q^{*}(R)$ shows that what 
we need to prove is that
\begin{eqnarray}
&\;& \min \{S(B|C):S(X:C)+S(B|C)=R+Q^{*}(R)\}\\
&\geq& \min\{S(B|C):S(X:C)=R\}.
\end{eqnarray} 
Let $ \omega $ be the state that minimizes the first expression for fixed $R$.
If $S(X:C)_\o \leq R$ then we're done so we may suppose not:
$S(X:C)_\o = R+\Delta$ for some $ \Delta > 0$. By convexity and
the definition of $H_c$, for any $R \geq H_c$,
\begin{equation}
 \frac{Q^{*}(R+\Delta)-Q^{*}(R)}{\Delta} > -1.
\end{equation}
Rearranging this inequality yields
\begin{equation}
 (R+\Delta) + Q^{*}(R+\Delta)  >   R+Q^{*}(R).
\end{equation}
Using the hypothesis $S(X:C)_\o = R+\D$ and the fact that the
right hand side of the above inequality is $S(X:C)_\o + S(B|C)_\o$,
we find that $S(B|C)_\o <  Q^{*}(R+\Delta)$. But, again by hypothesis,
$S(X:C)_\o=R + \D $ so we have a contradiction of the definition
of $Q^{*}(R+\D)$. We conclude that $ S(X:C)_\o \leq R$. 

Finally, $R+Q^*(R)-\bar{S} = \chi$ when $R < H_c$ so $E^*(R) = E^*(\chi)$
is constant. Using the first half of the lemma, we then find
$E^*(\chi) = E^*(H_c+Q^*(H_c)-\bar{S}) = Q^*(H_c)$.
\end{proof}

\begin{lemma} \label{lem:2}
$Q^{*}(R-E^{*}(R)+\bar{S}) = E^{*}(R)$ when $R \geq \chi$. 
Otherwise $E^*(R) = +\infty$.
\end{lemma}
\begin{proof}
Let $H_c \leq R_1$ and consider $R = R_1 + Q^*(R_1) - \bar{S}$.
$R$ is a strictly increasing function of $R_1$ by the definition of $H_c$,
taking all values $\chi \leq R$. Substituting into lemma~\ref{lem:1}
gives
\begin{eqnarray}
Q^{*}(R-E^{*}(R)+\bar{S}) 
& = & Q^{*}(R_1+Q^{*}(R_1)-\bar{S}-Q^{*}(R_1)+\bar{S})\\
& = & Q^{*}(R_1)\\
& = & E^{*}(R_1+Q^{*}(R_1)-\bar{S})\\
& = & E^{*}(R).
\end{eqnarray}
Also, $R < \chi$ is not achievable 
(by causality, see section \ref{subsec:forbidden}), yielding the 
second half of the lemma.
\end{proof}

\section{The triple trade-off} \label{main}

The following theorem is the main result of the paper: a prescription
for calculating the minimal amount of entanglement required given any
cbit and qubit rate.
\begin{theorem}\label{mainThm}

\[ E^{*}(R,Q)\;=\; \left \{ \begin{array}{ll}
			   0  & \mbox{if \; $Q^*(R) < Q$} \\
                           Q^{*}(R)-Q   &    \mbox{if $\;  \frac{1}{2}( Q^{*}(R)- \bar{S}) \leq Q \leq Q^{*}(R) $}  \\
                           E^{*}(R+2Q)-Q &  \mbox{if $\;  \frac{1}{2} (\chi-R) \leq Q < \frac{1}{2}( Q^{*}(R)- \bar{S}) $}\\
                           +\infty   &     \mbox{if$ \; Q <  \frac{1}{2} (\chi-R) $}
                           \end{array}
                    \right. \]
\end{theorem}
We discuss each of the four ranges for $Q$ separately, referring to them,
in order, as the \emph{QCT region}, the 
\emph{low-entanglement region}, the \emph{high-entanglement region} 
and the \emph{forbidden region}. The names of the first and last
regions should be self-explanatory. (QCT is optimal by definition in the
QCT region and no amount of entanglement is sufficient in the forbidden
region.) In the low-entanglement region
we'll find that optimal protocols can be found by time-sharing between
QCT and SDC (the first of which does not use entanglement) while the
optimal protocols for the high-entanglement region are found by
time-sharing between RSP and SDC, \emph{both} of which rely on entanglement.

While $H_c$ does not appear explicitly in our formula, it once again
delineates the boundary between two qualitatively different regimes: 
for $R<H_c$ we have that 
$\smfrac{1}{2}(Q^{*}(R)-\bar{S})=\smfrac{1}{2}(\chi-R)$ 
so there is no high-entanglement region in this case.
The region defined by $R<H_c$ and $Q\geq \smfrac{1}{2}(\chi-R)$ is 
entirely contained in low-entanglement region.

\begin{figure}[t]
\begin{center}
\epsfxsize=5.5in \epsfbox{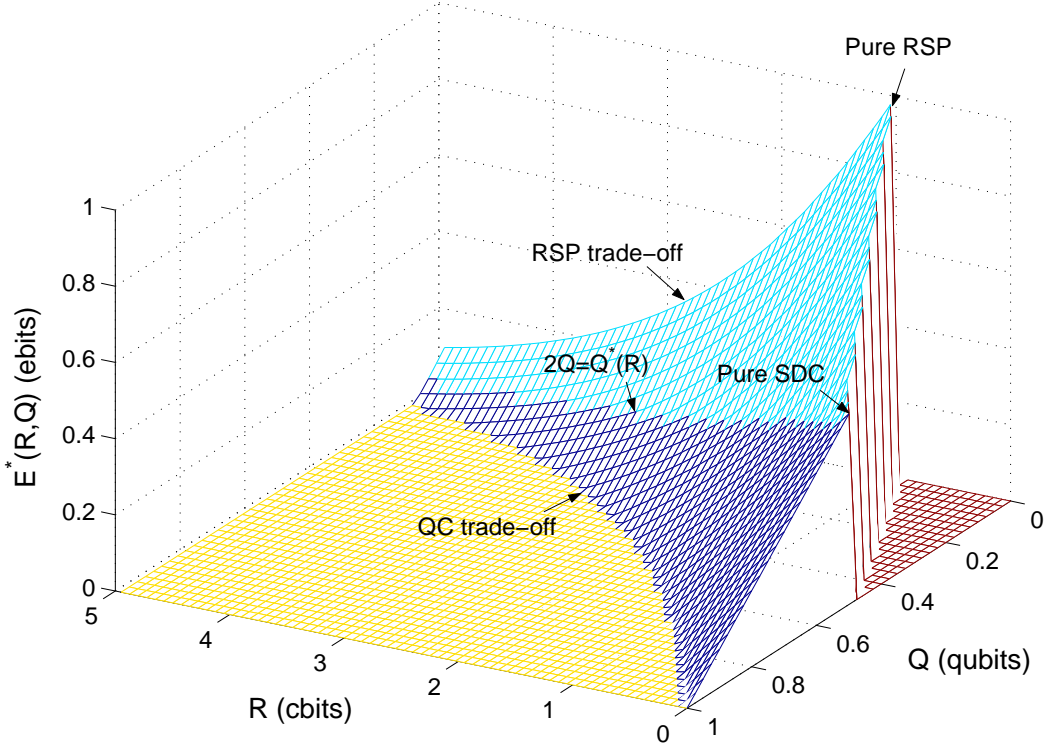}
\end{center}
\caption{Trade-off surface for the uniform qubit ensemble. The region
on the left for which $E^*(R,Q)=0$ is the QCT region, whose boundary with
the low-entanglement region is given by the curve $(R,Q^*(R),0)$. 
The transition to the high-entanglement region then occurs when
$2Q = Q^*(R)$; note that the surface is not smooth at the transition.
Finally the points corresponding to pure RSP, $(1,0,1)$, and pure
SDC, $(0,1/2,1/2)$, define the boundary of the forbidden region.
In the low-entanglement region, the trade-off is a ruled surface, 
linear for constant $R$.}
\label{fig:qubit}
\end{figure}

Before giving a proof of theorem \ref{mainThm}, we consider the
standard example: $\cE_{AB}$ being the uniform (unitarily invariant)
ensemble over qubit states on $B$. Devetak and Berger gave an explicit
parametrization~\cite{DB01} of the function identified as $Q^*(R)$ 
for this ensemble in Ref.~\cite{HJW02} and the
corresponding RSP curve appeared in Ref.~\cite{BHLSW03}. 
We present the full
trade-off surface $E^*(R,Q)$ in figure~\ref{fig:qubit}.
(In the case of 
an infinite ensemble, theorems \ref{thm:RSP} and \ref{thm:QCT} need to be
slightly modified: the $\min$ should be replaced by an $\inf$ as explained
in theorem 10.1 of Ref.~\cite{HJW02}. The only 
modification required to the argument
of this paper is in the second half of lemma \ref{lem:1}, where a
sequence of $\o_n$ needs to be considered instead of a fixed minimizing
$\o$.)

We also summarize for convenience in table \ref{table:triples} 
all the rate triples
and conversions between them that we will use in the proof. 
We use the notation $(R,Q,E) \longrightarrow (R',Q',E')$ to indicate that if 
the rate triple $(R,Q,E)$ is achievable 
then so is the rate triple $(R',Q',E')$;
\emph{i.e.} $(R,Q,E)$ can be \emph{converted}
into $(R',Q',E')$. Similarly, if we write 
$(R,Q,E)\optimal \lrar (R',Q',E')$ then the conversion is possible
conditional on $(R,Q,E)$ being optimal.

\begin{table}[t]
\begin{center}
\caption{Achievable rate triples and conversions}
\label{table:triples}
\begin{tabular}{|l|l|}
\hline
Rate triple & Description \\ [0.5ex]
\hline \hline
$(R,Q^{*}(R),0)$ 
	&  QCT  \\
\hline
$(R,0,E^{*}(R))$ 
	&  RSP \\
\hline
$(R,\smfrac{1}{2}(Q^{*}(R)-\bar{S}),\smfrac{1}{2}(Q^{*}(R)+\bar{S}))$ 
	& SDC on QCT: Eq.~(\ref{eqn:SDC}) \\
\hline
$(R+Q^{*}(R)-\bar{S},0,Q^{*}(R))$ for $R\geq H_c$ 
	& QCT to RSP: lemma~\ref{lem:1} \\
\hline
$(R-E^{*}(R)+\bar{S},E^{*}(R),0)$ 
	& RSP to QCT: lemma~\ref{lem:2} \\
\hline
$(R,Q,E)\longrightarrow(R+2Q,0,E+Q)$ 
	& Teleportation (of qubits) \\
\hline
$(R,Q,E)\longrightarrow
	(0,Q+\smfrac{1}{2}R+Q,\smfrac{1}{2}R+E)$ 
	& Superdense coding (of cbits) \\
\hline
$(R_1,Q_1,E_1) \;\&\; (R_2,Q_2,E_2)$ &  \\
	$\quad\quad\quad \longrightarrow \l (R_1,Q_1,E_1) 
	+ (1-\l) (R_2,Q_2,E_2)$ 
	& Time-sharing \\
\hline
$(R,Q,E) \longrightarrow (R,Q+E,0)$ 
	& Sending entanglement using qubits\\
\hline
$(R,Q,E)\optimal \longrightarrow (R-E+Q-\bar{S},Q+E,0)$ & \\
	\quad\quad\quad if $R \geq \bar{S}$ and $E > Q + \bar{S}$ 
	& Lemma~\ref{lem:4} \\
\hline
\end{tabular}
\end{center}
\end{table}

\subsection{ The low-entanglement region:  $\; \frac{1}{2}( Q^{*}(R)-\bar{S}) \leq Q \leq Q^{*}(R) $ }

Define $\l=2(Q^{*}(R)-Q)/(Q^{*}(R)+\bar{S})$. By the definition
of the low-entanglement region, $0 \leq \l \leq 1$. Both 
$(R, Q^{*}(R),0)$ and 
$(R, \frac{1}{2}(Q^{*}(R)-\bar{S}),\frac{1}{2}(Q^{*}(R)+\bar{S}) )$ 
are achievable so the convex combination
\begin{eqnarray}
(R,Q,Q^{*}(R)-Q) 
=  \l (R,Q^{*}(R),0)\;+\; 
(1-\l)\left(R,\frac{1}{2}(Q^{*}(R)-\bar{S}),\frac{1}{2}(Q^{*}(R)+\bar{S})\right)
\end{eqnarray}
is achievable by time-sharing.

The proof that these points are optimal is very simple. 
Suppose they are not. Then there 
would exist an $\epsilon$ such that  $  (R,Q,Q^{*}(R)-Q-\epsilon) $ were 
optimal. Now, using the conversion $(R,Q,E)\rightarrow (R,Q+E,0)$, 
it follows that 
$ (R,Q^{*}(R)-\epsilon,0) $ is achievable, which is a contradiction
of the definition of $Q^*$.

\subsection{ The high-entanglement region:  $\frac{1}{2} (\chi-R) \leq Q < \frac{1}{2}( Q^{*}(R)-\bar{S})$}

This region seems to require a more elaborate analysis. 
We first define two new variables $R_1$ and $R_2$ which are 
functions of $R$ and $Q$ but much easier to work with: 
\begin{eqnarray}
 R_1 &=& R+2Q-E^{*}(R+2Q)+\bar{S} \\
 R_2 &=& R-R_1+\bar{S}\;=\;E^{*}(R+2Q)-2Q.
\end{eqnarray}
We collect for future use some simple facts about $R_1$ and $R_2$:
\begin{enumerate}
\item \label{fact:0} $R_1 \geq H_c \; :$ \\
The function $R' - E^*(R') + \bar{S}$ is a monotonically increasing
function of $R'$. By causality, therefore, the minimum of this function
over achievable $R'$ occurs when $R' = \chi$. From lemma \ref{lem:1},
$E^*(\chi) = Q^*(H_c) = S - H_c$, so $R' - E^*(R') + \bar{S} \geq H_c$.\
Since $R+2Q \geq \chi$ in the high-entanglement region, we conclude that
$R_1 \geq H_c$.
\item  \label{fact:1} 
$ Q=\frac{1}{2}(Q^{*}(R_1)-R_2) \; :$ \\
This follows by lemma~\ref{lem:2}: $ Q^{*}(R_1)\;=\; E^{*}(R+2Q)\;=\; R_2+2Q$.
\item \label{fact:2} 
$E^{*}(R+2Q)-Q \;=\;R_2+Q \;=\;\frac{1}{2}(Q^{*}(R_1)+R_2) \; :$ \\
This follows by the  definition of $R_2$ and the previous fact.
\item \label{fact:3} 
$ R_2 \leq Q^{*}(R_1) \; :$ \\
By fact 1, $R_2=Q^{*}(R_1)-2Q $.
\item \label{fact:4} 
$Q^{*}(R_1) \geq \bar{S} \; :$ \\
$ Q^{*}(R_1)-\bar{S}
	\;=\; S(B|C)-\bar{S}
	\;=\;S(X:B|C)\geq 0$ (for optimal $\o$).
\item \label{fact:5}
$R_2 \geq \bar{S}$ 
	(for $Q \leq \frac{1}{2}( Q^{*}(R)-\bar{S})$) $\; :$ \\
This is equivalent to $ E^{*}(R+2Q) \geq 2Q +\bar{S} $. 
Since $2Q \leq Q^{*}(R)-\bar{S}$ in this region, we have by 
the monotonicity of $E^{*}$ and by lemma~\ref{lem:1} that
\begin{eqnarray}
  E^{*}(R+2Q) & \geq & E^{*}(R+Q^{*}(R)-\bar{S})\\
              &  =   & Q^{*}(R)\\
              & \geq & 2Q+\bar{S}.
\end{eqnarray}
\end{enumerate}
Equipped with these observations we can now proceed to the proof of 
theorem \ref{mainThm} in the high-entanglement region. That is, we will
prove that
$ E^{*}(R,Q) = E^{*}(R+2Q)-Q$ when
$\smfrac{1}{2}(\chi -R) \leq Q < \frac{1}{2}(Q^{*}(R)-\bar{S})$.
Note that 
\begin{equation} \label{eqn:highEntanglement}
\left(R,Q,E^{*}(R+2Q)-Q\right) = 
\left(R_1+R_2-\bar{S}, \smfrac{1}{2}(Q^{*}(R_1)-R_2),
\smfrac{1}{2}(Q^{*}(R_1)+R_2) \right)
\end{equation} 
in terms of the new variables, by the definition of $R_1$ and $R_2$ 
as well as facts \ref{fact:1} and 
\ref{fact:2}.

\subsubsection{Proof of achievability}

$(R_1, \frac{1}{2}(Q^{*}(R_1)-\bar{S}), 
\frac{1}{2}(Q^{*}(R_1)+\bar{S}) ) $ 
is achievable by Eq.~(\ref{eqn:SDC}) and 
$(R_1+Q^{*}(R_1)-\bar{S},0,Q^{*}(R_1)) $ is achievable by 
lemma~\ref{lem:1}. By facts \ref{fact:3},\ref{fact:4},and \ref{fact:5},
$\l = (Q^*(R_1) - R_2)/(Q^*(R_1)-\bar{S})$ is between $0$ and $1$.
Therefore, the convex combination
\begin{eqnarray}
&\;& \left(R_1+R_2-\bar{S}, \smfrac{1}{2}(Q^{*}(R_1)-R_2),
	\smfrac{1}{2}(Q^{*}(R_1)+R_2)\right) \\
&=& \l \left(R_1, \smfrac{1}{2}(Q^{*}(R_1)-\bar{S}), 
	\smfrac{1}{2}(Q^{*}(R_1)+\bar{S}) \right) +
	(1-\l) \left(R_1+Q^{*}(R_1)-\bar{S},0,Q^{*}(R_1)\right)
\end{eqnarray}
is also achievable by time-sharing.

\subsubsection{Proof of optimality}

We defer the proof of the following lemma, which is at the heart of our
optimality proof, to the end of the section:
\begin{lemma} \label{lem:4}
If $R_1, Q \geq 0$ and $R_2 > \bar{S}$, then there is a conversion
\begin{equation}
(R_1+R_2,Q,R_2+Q)\optimal \longrightarrow (R_1+\bar{S},R_2+2Q,0). 
\end{equation}
\end{lemma}
(Note that when $R_2 = \bar{S}$, the conversion always exists, regardless
of the optimality of the first rate triple.)
Now suppose that points of the form of Eq.~(\ref{eqn:highEntanglement}) are
not optimal. Then there exists some $\epsilon > 0 $ such that 
\begin{equation} \label{eqn:optimal2}
\left (R_1+R_2-\bar{S}, \smfrac{1}{2}(Q^{*}(R_1)-R_2),\smfrac{1}{2}(Q^{*}(R_1)+R_2)-\epsilon \right) 
\end{equation}
is optimal. We handle the cases $R_2 > \bar{S} + \e$ and 
$R_2 \leq \bar{S} + \e$ separately.

Assume first that $R_2 > \bar{S} + \e$, then
define $R_1^{\prime} = R_1-\bar{S}+\epsilon$ and 
$R_2^{\prime} =  R_2-\epsilon$.
Rewriting the triple~(\ref{eqn:optimal2}) in terms of 
$R_1^{\prime}$ and $R_2^{\prime}$, we have that
\begin{equation}
 \left(R_1^{\prime}+R_2^{\prime}, \smfrac{1}{2}(Q^{*}(R_1)-R_2), R_2^{\prime} + \smfrac{1}{2}(Q^{*}(R_1)-R_2)\right )
\end{equation}
is optimal.
Since $R_2^{\prime}>\bar{S}$, we can use lemma~\ref{lem:4} to obtain that
$(R_1+\epsilon,Q^{*}(R_1)-\epsilon,0)$ is achievable. 
This implies that $ Q^{*}(R_1+\epsilon) \leq Q^{*}(R_1)-\epsilon $, 
which is  a contradiction since, by fact \ref{fact:0}, $R_1 \geq H_c $.

If instead $R_2 \leq \bar{S}+\epsilon$,
we apply the conversion $(R,Q,E) \longrightarrow (R,Q+E,0)$ 
obtained by using quantum communication to establish entanglement:
\begin{equation}
 \left(R_1+R_2-\bar{S}, \smfrac{1}{2}(Q^{*}(R_1)-R_2),\smfrac{1}{2}(Q^{*}(R_1)+R_2)-\epsilon \right)\optimal \longrightarrow \left(R_1+R_2-\bar{S},Q^{*}(R_1)-\epsilon,0 \right).
\end{equation}
This implies that $ Q^{*}(R_1+R_2-\bar{S}) \leq Q^{*}(R_1)-\epsilon $. 
We also have $Q^{*}(R_1+\epsilon) \leq Q^{*}(R_1+R_2-\bar{S})$
by assumption and the monotonicity of $Q^{*}$. As before, we find
that $Q^{*}(R_1+\epsilon) \leq Q^{*}(R_1)-\epsilon $, which is  
a contradiction.

\vspace{2mm}

\begin{proof}\proofComment{Of lemma \ref{lem:4}}
Performing teleportation yields the conversion
\begin{equation}
(R_1+R_2,Q,R_2+Q)\longrightarrow (R_1+R_2+2Q,0,R_2+2Q).
\end{equation} 
(Note that teleportation is appropriate here instead of RSP because
the encoding map corresponding to the first triple will generally
produce complicated entangled states between Alice and Bob, conditioned 
on the classical bits being communicated. Teleportation will preserve
this entanglement.)
It will suffice to prove that the resulting triple is optimal because
an application of lemma \ref{lem:2} would then show that
$(R_1+\bar{S},R_2+2Q,0)$ is achievable.

Suppose then that $(R_1+R_2+2Q,0,R_2+2Q)$ is not optimal so that
there exists some $ \epsilon > 0 $ such that $(R_1+R_2+2Q,0,R_2+2Q-\epsilon)$ 
is optimal. By lemma \ref{lem:2} and then Eq.~(\ref{eqn:SDC}), 
there is a sequence of conversions
\begin{eqnarray} \label{eqn:TPoptimal}
&\;& (R_1+R_2+2Q,0,R_2+2Q-\epsilon) \optimal \\
&\longrightarrow& (R_1+\epsilon+\bar{S},R_2+2Q-\epsilon,0)\optimal 
	\label{eqn:step1} \\
&\longrightarrow& \left(R_1+\e+\bar{S},\smfrac{1}{2}(R_2+2Q-\e-\bar{S}), 
	\smfrac{1}{2}(R_2+2Q-\e+\bar{S})\right)
\end{eqnarray}
We handle the cases $R_2 \geq \bar{S} + \e$ and $R_2 < \bar{S} + \e$
separately.

Assume first that $R_2 \geq \bar{S} + \epsilon $. Then if we
define $\l = (R_2 - \bar{S} - \e)/(R_2 + 2Q - \e - \bar{S})$, we have
$0 \leq \l \leq 1$ so the convex combination
\begin{eqnarray}
&\;& (R_1 + R_2, Q, R_2 + Q - \e) \\
&=& \l (R_1+R_2+2Q,0,R_2+2Q-\e) \\
&\;& \quad	+ (1-\l) \left( R_1 + \e + \bar{S},
		\smfrac{1}{2}(R_2+2Q-\e-\bar{S}), 
		\smfrac{1}{2}(R_2+2Q-\e+\bar{S})\right)
\end{eqnarray}
is achievable,
contradicting the optimality of $(R_1+R_2,Q,R_2+Q)$.

Now suppose that $R_2 < \bar{S} + \e$ and consider $\a = \e + \bar{S} - R_2$,
which is by definition positive. Rewriting the triple (\ref{eqn:step1})
in terms of $\a$, applying the SDC conversion of Eq.~(\ref{eqn:SDC}) and
then regular superdense coding of the cbits gives
\begin{eqnarray}
&\;& (R_1 + R_2 + \a, 2Q - \a + \bar{S}, 0)^* \\
&\longrightarrow& (R_1 + R_2 + \a, Q - \a/2, Q-\a/2 + \bar{S}) \\
&\longrightarrow& \left( 0, Q + \smfrac{1}{2}(R_1+R_2),Q
				+ \smfrac{1}{2}(R_1+R_2)+\bar{S} \right).
\end{eqnarray}
Choosing $\l = \a / (R_1 + R_2 + \a)$, we can time-share to achieve
\begin{eqnarray}
&\;& (R_1+R_2,Q,Q+\bar{S}) \\
&=& \l \left( 0, Q + \smfrac{1}{2}(R_1+R_2),Q
				+ \smfrac{1}{2}(R_1+R_2)+\bar{S} \right) \\
&\;& \quad	+ (1-\l) (R_1 + R_2 + \a, Q - \a/2, Q-\a/2 + \bar{S}),
\end{eqnarray}
contradicting again the optimality of $(R_1+R_2,Q,R_2+Q)$ since
$\bar{S} < R_2$ by the hypotheses of the lemma.
\end{proof}

\subsection{The forbidden region: $ Q < \frac{1}{2}(\chi-R) $ } 
\label{subsec:forbidden}

In keeping with the operational spirit of the other arguments in this paper,
we argue that achievability in this region would lead to a violation of
causality. A classical channel of dimension $d_C$ and a quantum
channel of dimension $d_Q$ can be used to transmit at most 
$\log d_C + 2 \log d_Q$ bits of classical information by the optimality of
superdense coding~\cite{BW92,H73}. Success in the ensemble communication
task, however, results in Bob holding a high-fidelity copy of $\cE_B$.
By using coding, Alice could then about communicate $\chi(\cE_B)$ classical
bits to Bob per usage of the protocol~\cite{SW97,Holevo98b}, a violation of 
causality (for sufficiently
high fidelity and small $\d$ in the notation of section \ref{sec:defn})
if $\chi(\cE_B) > R + 2Q$.

A simple entropic argument is also possible.
Consider the state 
\begin{equation}
\rho = \sum_{i^{n},j} p_{i^n} \proj{i^n}^X 
	\ox \r_{i^n,j}^{A B_1 B_2} \ox q(j|i^n) \proj{j}^C,
\end{equation}
which represents the output of Alice's encoding operation for a given 
(unspecified) protocol of the form of figure \ref{fig:circuit}. We can 
estimate
\begin{eqnarray}
\smfrac{1}{n} \chi( \{ \tilde{\ph}_{i^n}^B, p_{i^n} \} )
&\leq& S(X:B_1 \, B_2 \, C ) \quad \mbox{(by monotonicity of $\chi$)} \\
&=& S(X:B_2) + S(X:C|B_2) + S(X:B_1|B_2 C) \\
&\leq& \log d_C + 2 \log d_Q,
\end{eqnarray}
using the lemma \ref{lem:infoBound} (see below) 
twice and the fact that $S(X:B_2)=0$ since
$B_2$ is maximally mixed for all $i^n$. On the other hand, applying
the Fannes inequality~\cite{F73} and the fidelity condition implies that
\begin{equation}
\smfrac{1}{n} \chi( \{ \tilde{\ph}_{i^n}^B, p_{i^n} \} )
\stackrel{\e \rar 0}{\longrightarrow} \chi,
\end{equation}
giving the constraint $\chi \leq R + 2Q$. 

\begin{lemma} \label{lem:infoBound}
Let $\rho$ be a tripartite density operator of the form
\begin{equation}
\rho = \sum_i p_i \proj{i}^X \ox \rho_i^{AB},
\end{equation}
where the states $\{ \ket{i}^X \}$ are orthonormal and the $p_i$ are
probabilities. Then 
\begin{equation}
S(X:A|B) \leq \min( \log \dim X, 2 \log \dim A ).
\end{equation}
\end{lemma}
\begin{proof}
We can expand $S(X:A|B) = S(X|B) - S(X|AB)$. By 
subadditivity of the von Neumann entropy, 
the first term is less than or equal to $S(X)$,
which is in turn no more than $\log \dim X$.
Moreover, because $\r$ is separable across the $X/AB$ cut, $S(X|AB) \geq 0$.
(This follows immediately from concavity of the entropy~\cite{CA99,HHH98}.)

To prove the second inequality, we expand the definition of $S(X:A|B)$
differently:
\begin{equation}
S(X:A|B) 
= S(A|B)_{\r^{AB}} +  \sum_i p_i S(A|B)_{\r_i^{AB}}.
\end{equation}
Using subadditivity of the von Neumann entropy again, $S(A|B) \leq S(A)$ for
any density operator. $S(A)$, in turn, is always less than or equal to
$\log \dim A$. 
\end{proof}

\section{Discussion}
The problem we posed here, communication using noiseless classical
and quantum channels in addition to maximally entangled states, is the
natural setting in which to unify many pre-existing results
on quantum-classical compression, remote state preparation and quantum
state superdense coding. While our goal was to provide a unified synthesis
of these disparate results, our conclusion was ultimately that the 
the general problem can be understood in terms of those basic building
blocks -- the surface of optimal rate triples for the triple resource
problem can be assembled by time-sharing appropriately between protocols
designed for the special cases. Such a neat resolution confirms the
simplifying power of the resource-based approach and justifies viewing
trade-off coding, remote state preparation and quantum state
superdense coding
as fundamental primitives \emph{instead} of special
cases of a more general problem.

\subsection*{Acknowledgments}
We thank Debbie Leung for many helpful conversations and her
patience when faced with our invasions of her office space.
The authors acknowledge the support of the US National
Science Foundation under grant no. EIA-0086038. PH is also
supported by the Sherman Fairchild Foundation.

\bibliographystyle{unsrt}
\bibliography{triple}

\end{document}

%% file: circuit.eepic
\setlength{\unitlength}{0.000513333in}
\begingroup\makeatletter\ifx\SetFigFont\undefined%
\gdef\SetFigFont#1#2#3#4#5{%
  \reset@font\fontsize{#1}{#2pt}%
  \fontfamily{#3}\fontseries{#4}\fontshape{#5}%
  \selectfont}%
\fi\endgroup%
{\renewcommand{\dashlinestretch}{30}
\begin{picture}(10374,4221)(0,-10)
\put(312,3987){\ellipse{424}{424}}
\path(312,3762)(312,3237)(312,3312)
\path(312,3537)(12,3387)
\path(312,3537)(612,3387)
\path(312,3237)(87,2937)
\path(87,2937)(462,2937)
\path(312,3237)(537,2937)
\path(462,2937)(537,2937)
\path(237,2937)(237,2637)
\path(387,2937)(387,2637)
\put(10062,1362){\ellipse{424}{424}}
\path(10062,1137)(10062,387)
\path(10062,387)(9762,12)
\path(10062,387)(10287,12)
\path(10062,837)(9762,612)
\path(10062,837)(10362,612)
\path(9300,4100)(9400,4100)(9400,800)(9300,800)
\thicklines
\path(3237,4137)(3912,4137)(3912,3237)
	(3237,3237)(3237,4137)
\path(7317,1362)(7962,1362)(7962,462)
	(7317,462)(7317,1362)
\path(3237,3387)(2562,3387)(1212,1962)
	(2562,612)(7287,612)
\path(3912,3387)(4662,3387)(6612,912)(7287,912)
\blacken\path(7047.000,852.000)(7287.000,912.000)(7047.000,972.000)(7047.000,852.000)
\path(3912,3987)(9237,3987)
\path(7962,912)(9237,912)
\dashline{90.000}(3237,3987)(1212,3987)
\dashline{60.000}(3912,3687)(4662,3687)(6612,1212)(7287,1212)
\blacken\path(7047.000,1152.000)(7287.000,1212.000)(7047.000,1272.000)(7047.000,1152.000)
\put(3392,3612){$E_{i^n}$}
\put(7452,837){$D_j$}
\put(5262,2037){$B_1$}
\put(1212,4062){$X: \; i^n$}
\put(5262,707){$B_2$}
\put(9012,4062){$A$}
\put(9087,987){$B$}
\put(807,1887){$|\Phi\rangle$}
\put(5787,2412){$C:\;j$}
\put(9500,2500){$\tilde{\varphi}_{i^n}$}
\end{picture}
}